\documentstyle[11pt]{article}

\begin{document}
%\baselineskip 25pt
%\begin{titlepage}
\begin{flushright}
October 14, 1995
\end{flushright}

\vskip 10pt
\begin{center}
\Large\bf
Are the singularities stable?

\vskip 26pt

\normalsize

Peter K. Silaev \footnote{Electronic address:
silaev@sunny.bog.msu.su }  \\

\it Department of Physics and \\
Bogolyubov Institute for
Theoretical Microphysics, \\

Moscow State
University, Moscow, 119899, Russia.\\

\rm  and  \\

Slava G. Turyshev \footnote{ On leave from Bogolyubov Institute for
Theoretical Microphysics,

\hskip 8pt Moscow State University, Moscow, 119899, Russia.

\hskip 8pt Electronic address: sgt@zeus.jpl.nasa.gov }  \\

\it Jet Propulsion Laboratory MS 301-230, \\
California Institute of Technology \\
 4800 Oak Grove Drive,
Pasadena, CA 91109, USA.

\rm

\vskip 1pt

\end{center}

\begin{abstract}
The spacetime singularities play a useful role in gravitational theories
by distinguishing physical solutions from non-physical ones.
The problem,  we studying in this paper is: are these singularities
stable?
To answer this question, we have analyzed the
general problem of stability of the family of the static spherically
symmetric solutions of the standard
Einstein-Maxwell model coupled to  an extra free  massless scalar field.
We have obtained the equations for the axial and polar perturbations.
The stability against  axial perturbations has
been proven.
\vskip 15pt
\noindent PACS number(s): \hskip 2pt 04.20.Cv,
  04.20.Jb,   04.40.Nr,   04.80.Cc

\end{abstract}
%\end{titlepage}

\section{Introduction}

Recently there has been considerable interest in so-called ``dilaton fields'',
{\it i.e.} neutral scalar fields whose background values determine
the strength of the coupling constants in the effective
four-dimensional theory. However, although the
 scalar field naturally
arises in theory, its existence from the
point of view of the general relativity is quite problematic.
It has been shown that including a scalar field in
the theory   leads to a violation of the strong equivalence
principle and  modification of large-scale gravitational
phenomena \cite{DP}. The presence of the scalar field
affects  the equations of motion of the
other matter fields as well. Thus, for example, solutions which
correspond to a pure electromagnetic field appear to be
drastically
modified by the scalar field. Such solutions were studied in
\cite{Garfinke}-\cite{Wilczek}, where it was shown that the scalar field
 generally destroys the horizons leading to the singularities
 in a scalar curvature on a finite radii.
Special attention has been paid to the charged dilaton black hole
solution \cite{Garfinke}. Thus analysis of the perturbations around
the extreme charged dilaton black hole solution performed in \cite{Wilczek}
demonstrates the analogy of the behavior of the  black holes
and elementary particles in the sense that there exists an
energy gap in the excitation spectrum of the black hole.

 From the other side, an interesting way to treat the problem of appearance
of the spacetime singularities  is to develop a theory of gravity including
an extra spatial dimensions \cite{Witten}.
It turns out that certain singularities can be resolved  by simply passing
into a higher-dimentional theory of gravity for which spacetime is
only effectively four-dimensional below some
compactification scale \cite{GHT}.
 Moreover, while studying the decay of magnetic fields
in Kaluza-Klein theory was argued in \cite{Fay}
that for a physical four-dimensional
magnetic field  there are two ways it can decay: either
by producing single naked
singularities into which space ``collapses,'' or by producing pairs of
monopole-anti-monopole pairs which accelerate off to infinity.
Since many currently popular unified fields theories include an extra
spatial dimensions, it is important to ask: could  these singularities
be stable in our four-dimentional world?
Although, it was shown that static
spherically symmetric solutions for the related case
of Einstein-Klein-Gordon equations
with a quadratic self-interaction term  are unstable \cite{Jetzer},
it would be interesting to study this problem for the general case of the
Einstein-Maxwell-scalar system.

In this paper we consider the   problem of   stability
  of the general  class of static
spherically-symmetric  solutions of the standard
Einstein-Maxwell model with an extra free scalar field $\phi$
with  four-dimentional action taken to be:

$$ S =-{1 \over 16\pi}\int dx^4\sqrt{-g} \Big(  R -
2 \hskip 1pt g^{mn}\nabla_m \phi \nabla_n \phi + F^2\Big),\eqno(1) $$

\noindent where $F_{ab}=\nabla_{[a} A_{b]}$ is the usual
Maxwell field. The geometrical units $c=\gamma=1$ are used
through the paper as is the following
metric convention $(+---)$.
The  fields equations corresponding to action (1) are easily
calculated to be:

$$ R_{mn}= 2 \nabla_m\phi\nabla_n\phi   - 2 F_{mk}{F_n}^k +
{1\over 2}g_{mn}F^2,\eqno(2a)$$

$$ g^{ab}\nabla_a\nabla_b\phi=0, \qquad \nabla_a F^{ab}=0\eqno(2b) $$

The general static spherically symmetric solution to system
of equations (2) is well known \cite{Garfinke}-\cite{Silaev}
and it might be given by the following relations

$$ ds^2=u(r)dt^2-v(r)dr^2-w(r)d\Omega, \eqno(3a)$$

$$ v(r)={1\over u(r)}= q^2(r), \qquad w(r)=(r^2-\mu^2) q^2(r), \eqno(3b) $$

$$ \phi(r)={\phi_0\over 2\mu} \ln {r-\mu \over r+\mu },
\qquad A_0'(r)={Q\over w(r)},  \eqno(3c) $$

$$ q(r)=  p_{\mp}\Big({r-\mu\over r+\mu}\Big)^k +
p_{\pm}\Big({r+\mu\over r-\mu}\Big)^k, \eqno(3d) $$

$$ 2p_{\pm}= 1\pm \left(1+ {Q^2 \over 4\mu^2 k^2}\right)^{1/2}, \qquad
\phi_0= \mu \sqrt{ 1-4k^2}, \eqno(3e)  $$

\noindent where $\mu, k, Q$  are the arbitrary constants,
 the prime denotes the derivative $d/dr$ and the usual notation is accepted
in $(3a)$ for
$d\Omega=d\theta^2+\sin^2\theta d\varphi^2$.  The parameter  $\mu$
is related to physical mass $\mu_0>0$ and charge $Q$ by
$$\mu=\pm {1\over 2k} \sqrt{\mu_0^2-Q^2},$$
which saturates the bound $|Q|\le\mu_0$.
In the extreme limit $|Q|=\mu_0$, and the solution, independently on the
scalar field, accepts the familiar form of the extreme Reisner-Nordstr\"om
black hole solution:

$$ds^2=\left({1\mp{\mu_0\over R}}\right)^2dt^2-
\left(1\mp{\mu_0\over R}\right)^{-2}dR^2-R^2d\Omega,$$

$$ \phi(r)=0, \qquad R=r\pm\mu_0.$$

\noindent
In some special cases this solution coincides with well known
results which will  support our future conclusions. This
class of the solutions (3) describes the exterior
region of the black holes and the naked
singularities\footnote{In the special case $k=\pm1/2$ result (3)
 reduces to the Reisner-Nordstr\"om solution whose properties has been
 studied extensively \cite{Chandra}.}. It is important to ask:
could a distant observer
study the objects, located    under this spurious singularity at $r=\mu$?
The answer appears to be no. Indeed, let us imagine that the observer
will try to test this region using the perturbations of the fields involved.
Can the  infinite energy density (and  corresponding singularity
   in the equation for the perturbations) be an opaque boundary for the
   perturbations,  or there is a possibility that the  perturbations
might penetrate under the   surface $r=\mu$?
To answer this question, one might easily show that  $r=\mu$ is an
effectively infinite point in the case of Schwatzchild ($k=\pm1/2, Q=0$)
and Reisner-Nordstr\"om ($k=\pm1/2$) solutions. In the vicinity of this
surface there are solutions which propagate in both the
``in'' and ``out'' directions $(\sim e^{\pm \omega r^*})$  \cite{Chandra}.
Then, for a distant observer, the time of the fall  is logarithmically infinite
because   $g_{00}/g_{11}\sim (r-\mu)^2$
({\it i.e.} we have a horizon). This is true  even though the distance
to the horizon can be
traversed in finite proper time. In the general case of the solution (3)
with non-zero scalar field $(k\not= \pm 1/2  \hbox{ and } |Q| \not=\mu_0)$,
there are no ``in'' or ``out'' going waves and one can
see that because of the relation $g_{00}/g_{11}\sim (r-m)^d$
with $d<2$  the time of the fall is finite (it has no horizon).
 From the analysis presented in this paper  we will see that the
perturbations will ``stop'' at the point $r=\mu$ (the
singular point of the equation for perturbations) and thus the observer
will not be able to see the  singularity.
It is worth to note that, although the energy densities for both scalar and
electromagnetic fields in solution (3) are infinite when $r\rightarrow \mu$,
 one may show  by  strightforward calculation that the
energy (and the mass) of the solution remain
finite. This suggests that we may consider a small perturbations
around the solution (3) and linearize the field equations (2).
Moreover,  it is reasonable to expect that the
corresponding  energy of perturbations
will be small  (and therefore also finite)
as compared to the  energy of the solution (3).

In this paper we will study the general problem of  stability
of the  solution (3) which describes the ``exterior'' region
of the black holes and the naked singularities.
It is reasonable to note that one would not expect general
solutions with naked singularities to be stable since the total mass can
be negative. However, the analysis presented here will show that the
solution (3) is stable at least against  axial  perturbations which,
in the light of the results of \cite{Witten}, \cite{Fay}, makes this reaserch
specifically interesting for the general case
of Einstein-Maxwell-scalar system, superstrings and Kaluza-Klein theories.
The outline of this paper is as follows: In the next section  we will
introduce the definitions accepted throughout the paper and will
obtain the system of   equations for  axial and polar perturbations.
In section 3 we will study the problem of stability of the solution (3)
against   axial perturbations.
In the following section 4 we will examine the problem of   splitting
of the obtained $2\times 2$ matrix equation into two independent
equations. In the final section 5, we will
summarize our results and suggest the  perspectives of the research on
the problem of stability  of the static spherically symmetric solutions of the
Einstein-Maxwell-scalar system. We will also discuss  some possible
experimental consequences of  presence of the electromagnetic and
scalar fields in the motion of the celestial bodies.

\section{ The general system of the equations for the fields
perturbations }
\hskip 10pt

It is well-known that in the presence of the non-zero background matter
field, the perturbations of matter and gravitational fields should be
studied simultaneously. Otherwise, the equations of motion can appear
to be inconsistent  and, in any case, they can not be applied to the
stability problem of the solution under consideration.

In this section we will
obtain the general system of the equations for both
axial and polar perturbations for the system of equations (2).
In order to simplify the future calculations, let us introduce
notation  for the perturbations of scalar, electromagnetic and
gravitational fields. Due to the  symmetries of the background field, this can
be done at a rather straightforward way \cite{Regge}.  Indeed, as far as
the background field does not depend on time, we can write
the perturbation of any component $  f(t, \vec{x}) $
for any field involved as:

$$ \delta f(t, \vec{x}) = \exp(i \omega t) \delta f(\vec{x}). $$

\noindent Because of the spherical symmetry of the
background field we, following  \cite{Goldberg}, will define the
spin-weighted spherical harmonics by the  equations:

$$ \big(-\partial_2\pm s \cot\theta \mp {i \over \sin\theta }
\partial_3 \big) {}_{s}Y_{lm}(\theta,\varphi) = $$
$$= \sqrt{ (l \pm s + 1
) (l\mp s) } \hskip 2pt {}_{s\pm 1}Y_{lm}(\theta,\varphi), $$
\noindent where
$$ {}_{0}Y_{lm}(\theta,\varphi) = Y_{lm}(\theta,\varphi). $$

\noindent Then we may expand all the spin-weighted
perturbations through these spherical harmonics as follows:

(i). The "scalar" perturbations (i.e. the perturbations of the components
without the angular indices) can be given by:

$$ \delta \phi(\vec{x}) = \sum_{lm} z_{lm}(r) Y_{lm}(\theta,\varphi),
\eqno(5a) $$

$$ \delta A_0(\vec{x}) = \sum_{lm} k_{lm}(r)  Y_{lm}(\theta,\varphi), \qquad
\delta A_1(\vec{x}) = \sum_{lm} n_{lm}(r)  Y_{lm}(\theta,\varphi), \eqno(5b)$$

$$ \delta g_{00}(\vec{x}) = \sum_{lm} a_{lm}(r)  Y_{lm}(\theta,\varphi),
\qquad \delta g_{01}(\vec{x}) = \sum_{lm} b_{lm}(r)  Y_{lm}(\theta,\varphi),$$

$$\delta g_{11}(\vec{x}) = \sum_{lm} c_{lm}(r)
 Y_{lm}(\theta,\varphi).\eqno(5c) $$

(ii). The "vector" perturbations (i.e. the perturbations of the
components with one angular index only) can be expanded  \cite{Goldberg}
with respect to the spin-weighted spherical harmonics with the
spin $\pm 1$ as:

$$ \delta g_{02}(\vec{x}) \pm {i \over \sin\theta} \delta g_{03}(\vec{x}) =$$
$$ =
\sum_{lm} -l(l+1) \big[d_{lm}(r) \mp i e_{lm}(r)\big] {}_{\pm 1}Y_{lm}
(\theta, \varphi), \eqno(6a)$$

$$ \delta g_{12}(\vec{x}) \pm {i \over \sin\theta} \delta g_{13}(\vec{x}) =$$
$$=
 \sum_{lm} -l(l+1) \big[f_{lm}(r) \mp i g_{lm}(r)\big] {}_{\pm
1}Y_{lm}(\theta,\varphi), \eqno(6b)$$

$$ \delta A_2(\vec{x}) \pm {i \over \sin\theta} \delta A_3(\vec{x}) = $$
$$=\sum_{lm}-l(l+1) \big[o_{lm}(r) \mp i s_{lm}(r)\big] {}_{\pm 1}
Y_{lm}(\theta,\varphi).\eqno(6c) $$

(iii). And, finally, the "tensor" perturbations can be expanded \cite{Goldberg}
with respect to the spin-weighted spherical harmonics with the
spin $\pm 2, 0$ as:

$$ \delta g_{22}(\vec{x}) +
{1 \over \sin^2 \theta} \delta g_{33}(\vec{x})  = $$
$$=\sum_{lm} \big[h_{lm}(r) - l(l+1) \epsilon_{lm}(r)\big]
{}_{\pm 2}Y_{lm} (\theta,\varphi), \eqno(7a)$$

$$ \delta g_{22}(\vec{x})  - {1 \over \sin^2 \theta} \delta g_{33}(\vec{x})
 \pm {2i \over \sin \theta} \delta g_{23}(\vec{x})  = $$
$$= \sum_{lm}
\sqrt{(l-1)l(l+1)(l+2) } \big[\epsilon_{lm}(r) \mp 2i j_{lm}(r)\big]
{}_{\pm 2}Y_{lm}(\theta,\varphi).\eqno(7b) $$

In order to reduce the effective number of the variables we will
perform the  gauge transformation:

$$ x_a \to x_a + \xi_a,\eqno(8a) $$

\noindent where the components of the four-vector $\xi_a(t, \vec{x})$ are
given by the relations:

$$ \xi_0(t, \vec{x}) = \sum_{lm} \alpha_{lm}(r) Y_{lm}(\theta,\varphi),$$
$$ \xi_1(t, \vec{x})
 = \sum_{lm} \beta_{lm}(r) Y_{lm}(\theta,\varphi), $$
{}
$$ \xi_2(t, \vec{x}) \pm {i \over \sin\theta} \xi_3(t, \vec{x}) = $$
$$=\sum_{lm}-l(l+1)
\big[\gamma_{lm}(r) \mp i \delta_{lm}(r)\big] {}_{\pm 1}
Y_{lm}(\theta,\varphi).\eqno(8b)$$

\noindent We will impose the same conditions on the
 coefficients as in \cite{Regge}:

$$ \gamma_{lm}={1\over2} i_{lm} , \qquad \delta_{lm} = j_{lm} ,$$
$$ \alpha_{lm} =d_{lm} - i\omega \gamma_{lm}, \qquad \beta_{lm}
=f_{lm} - \gamma'_{lm} + \gamma_{lm}{w'\over w}. \eqno(9)$$

\noindent
Furthermore, we will introduce an additional set of   convenient
notations (with tildas) given by the following relations:

$$ a_{lm}=\tilde a_{lm} + 2 i \omega \alpha_{lm} - \beta_{lm}
{u'\over v},  $$

$$ b_{lm}=\tilde b_{lm} + i \omega \beta_{lm} +\alpha'_{lm} -
\alpha_{lm} {u'\over u},  $$

$$ c_{lm}=\tilde c_{lm} + 2 \beta'_{lm} - \beta_{lm} {v'\over v},$$

$$ b_{lm}=\tilde b_{lm} + 2 i \omega \alpha_{lm} - \beta_{lm}
{u'\over v}, $$

$$ e_{lm}=\tilde e_{lm} + i \omega \delta_{lm},  $$

$$ g_{lm}=\tilde g_{lm} + \delta'_{lm} - \delta_{lm} {w'\over w}, $$

$$ h_{lm}=\tilde h_{lm} + \beta_{lm} {w'\over v}, $$

$$ z_{lm}=\tilde z_{lm} - {\phi'\over 2 v}\Big(2f_{lm}-\epsilon'_{lm} +
\epsilon_{lm}{ w' \over w}\Big). \eqno(10) $$

The notations introduced above   significantly simplify the
future analysis of the perturbations of the equations of motion.
Thus, by  expanding the equations of motion (2) over the field
variations and then separating   the terms with  different angular
dependence (i.e. terms, proportional to ${}_0Y_{lm}(\theta,\varphi)$,
\ ${}_{\pm 1}Y_{lm}(\theta,\varphi)$, $\ldots$), one can easily find the
correspondent equations for the perturbations.
In particular, from expressions for the components $R_{22}$, $R_{23}$ and
$R_{33}$ given by the equations (2a) we will obtain the
following relations:

$$ \tilde g'_{lm}= \tilde g_{lm} \left({v' \over v} - {u' \over u}\right)+
i\omega{v \over u} \tilde e_{lm}. \eqno (11a) $$

\noindent Another equation might be obtained from the
expressions  for the components $R_{12}$ and $R_{13}$ $(2a)$, namely;

$$ \tilde e'_{lm}=\tilde e_{lm} {w' \over w} +
 \tilde g_{lm}{u \over i\omega w }\left(l(l+1)-2-\omega^2{w \over u }\right) +
 2  \tilde s_{lm} A'_0.  \eqno (11b) $$

\noindent And finally from the second  equation in $(2b)$ one may find
the last equation:

$$\tilde s''_{lm} =
\tilde s'_{lm}\left( { v' \over 2 v }- { u' \over 2 u }\right) +  \eqno (11c)
$$
$$+ \tilde s_{lm}{ v \over w }\left( l(l+1)  - \omega^2 {w\over u }\right) +
 \left(\tilde e'_{lm}  - \tilde e_{lm} {w'\over w } -
i\omega \tilde g_{lm}\right)
{ A'_0 \over u }.$$

Thus,  we have obtained three independent components of the
perturbations. These components are interacting only with each other
\cite{Regge} and hence they have no influence on the other components.
This is the trivial consequence of the fact  that these components
are axial, i.e.  when the spatial coordinates are inverted, their
transformation rules appears to be $-(-)^l$ rather then $(-)^l$.

Analogously, the general system for the polar perturbations takes the
form:

$$ \tilde a_{lm}'=  \tilde a_{lm} \left({ u' \over 2 u }+
{ w' \over 2 w}\right)+ \tilde b_{lm}{u\over i {\omega} w}
\left({l(l+1) \over  2}-\omega^2 {w\over  u}\right)- $$

$$- \tilde c_{lm} { u' \over 2 v } +\tilde h_{lm}
{ u \over w}\left({ u' \over 2 u}-{w' \over 2 w}\right) +
 \tilde k_{lm} 2 {A_0'}-  \tilde z_{lm}  u {\phi'},   $$

$$ \tilde b_{lm}'=  \tilde b_{lm} \left({ v' \over 2 v }-
{ u' \over  2 u}\right)  +  i {\omega}\left(\tilde c_{lm}  +
\tilde h_{lm} { v \over w }\right) + \tilde n_{lm}  2 {A_0'},  $$

$$ \tilde c_{lm} = \tilde a_{lm} { v \over u},   $$

$$ \tilde h_{lm}' = \tilde h_{lm} \left({ u' \over 2 u }+{w' \over 2 w}\right)
+ \tilde b_{lm} {l(l+1) \over 2i\omega} +
\tilde c_{lm} { w'\over 2 v}   +
\tilde z_{lm} w {\phi'},   $$

$$ \tilde k_{lm}'= \tilde a_{lm}  {  A_0' \over 2 u}  -
\tilde c_{lm} { A_0' \over 2 v}  + \tilde h_{lm} { A_0' \over w }
+ \tilde n_{lm}{u\over i {\omega}w}
\left( l(l+1) - \omega^2{ w \over  u }\right),   $$

$$ \tilde n_{lm}'= \tilde n_{lm} \left({ v' \over 2 v}- { u' \over 2 u}\right)+
\tilde k_{lm} {i \omega v \over u },   $$

$$ \tilde z_{lm}'=  - \tilde z_{lm} \left({w' \over  w } +
{u'\over  2u }\right)  -  $$

$$- { \tilde a_{lm} \over    u  \phi'}
\left[ \left({ l(l+1) \over 2} -1\right){v\over w}+
{ {A_0'}^2 \over 2u }+   { 3 u' w' \over  4 u w}\right] + $$

$$ + {\tilde h_{lm}  \over    w \phi'}\left[\left({ l(l+1) \over 2} -1\right)
{v\over w}+ {u' \over 2u}\left({w' \over 2w}-
{u' \over 2u}\right)+{ {A_0'}^2
\over u }-  \omega^2{v\over u} \right] + $$

$$ + {\tilde b_{lm}\over uw\phi'}{i\over 2\omega}
  \left( {l(l+1) \over 2}u' -\omega^2w'\right)
 +  \tilde k_{lm}{w' \over  w}
{ A_0' \over u \phi'}+
 \tilde n_{lm} {  l(l+1) A_0' \over  i \omega w \phi'}.  \eqno(12)$$

This is the system of independent equations for polar perturbations.
All the other equations appear to be a consequence of them.

\section{ The stability against the axial perturbations }

\hskip 12pt In this section we will concentrate on the stability
of the solution (3) against the  axial perturbations. In order to approach
this problem, we must rewrite the
system (11) as an eigenproblem with respect to $\omega^2$. There   exists
only one way to combine the first two equations of the system (11)  into
a single equation of the Shr\"oedinger-type  where
  $\omega^2$ playing the role of energy. To show this, let us define
the following combination:

$$ C_{lm} = \lambda_1 \tilde e_{lm} + \lambda_2 \tilde g_{lm}.  $$

\noindent It is strightforward to check that the equation for $ C_{lm}''$
 acquires the form of the Shr\"oedinger equation with that substitution
only when $\lambda_1=0$.

To present the system of equations for $\tilde g_{lm} $ and $\tilde s_{lm} $
in
the hermitian form, it convinient to introduce new functions $\psi_1$
and $\psi_2$ as follows:

$$ \tilde g_{lm} =\psi_1 w^{1/2} \Big({v\over u}\Big)^{3/4}  8i \omega,
\eqno(13a) $$

$$ \tilde s_{lm} = 4 \psi_2 \Big({v\over u}\Big)^{1/4} \sqrt{l(l+1)-2}.
\eqno(13b) $$

Then the equations for the column \hskip 2pt $\psi =
\left({ \psi_1 \atop \psi_2}\right) $ takes the form:

$$ \left({ \psi_1 \atop \psi_2}\right)''+ \omega^2 \rho(r) \left({
\psi_1 \atop \psi_2}\right) + \pmatrix{d(r)+a(r) &b(r) \cr b(r)& d(r) \cr }
 \left({ \psi_1 \atop \psi_2}\right) = 0, \eqno (14a) $$

\noindent where the   functions $a(r), b(r),
d(r)$ and weight $\rho(r)$ are defined form the relations (3) and
 given as follows:

$$ a(r)= {3\mu^2(1+4k^2)\over (r^2-\mu^2)^2}+
{6r\over r^2-\mu^2}  {q(r)'\over q(r)}, \eqno (14b) $$

$$ b(r)=- {2Q\over q(r)}{\sqrt{l(l+1)-2}\over (r^2-\mu^2)^{3/2}}, \eqno (14c)
$$

$$ d(r)=-{\mu^2(3+16k^2)\over (r^2-\mu^2)^2}- \eqno (14d)$$
$$- {6Q^2\over q(r)^2 (r^2-\mu^2)^2}
 - {l(l+1)\over (r^2-\mu^2)} - {8r\over r^2-\mu^2} {q'(r)\over q(r)}, $$

$$ \rho(r)=q(r)^4. \eqno (14e) $$

In order to prove the stability of the solution (3) with respect to
axial perturbations, following Wald  \cite{Wald} it is necessary  to show that:
 (i)
the spectrum of the differential operator (14) is positive, (ii) the
differential operator (14) is not only hermitian, but also
self-adjoint.

Concerning the eigenvalues of the operator (14), by
straightforward computation  one can  prove that for any
$\mu<r<\infty$ both eigenvalues of the potential matrix are positive
and then that the eigenvalues of the total operator (14) are also
positive.

Now let us analyze the self-adjointness of (14). The boundary
conditions at the spatial infinity ($r\to+\infty$) are fixed by the
means of the standard procedure:

$$ \int\!\!dr \, q(r)^4 \psi^+\psi < \infty \eqno(15) $$

\noindent and need no  further consideration. However, the
condition (15) with  $r\to \mu$ permits both possible asymptotics for
the function $\psi(r)$:

$$ \psi_1(r) = {\it const} \cdot (r-\mu)^{1/2\pm (s-1) }, \eqno(16a) $$

$$ \psi_2(r) = {\it const}\cdot (r-\mu)^{1/2\pm (s-1)/2 }, \eqno(16b) $$

\noindent where $s=|2k|$. By using the condition (15) one might immediately
conclude that because of the
relation:

$$ q(r\to \mu)={\it const} \cdot (r-\mu)^{-2s}. $$

\noindent the positive sign in (16a) is forbidden.
It means that in order to make the differential operator in (14)
  self-adjoint, we must impose some reasonable boundary condition  that
will suppress one of possible asymptotics of $\psi_2$ $(16b)$ for $r\to
\mu$.  The appropriate restriction appears to be quite natural: to impose
the condition of finiteness of the energy of electromagnetic perturbations. For
the
positive sign in the condition $(16b)$, the energy density is
proportional to $(r-\mu)^{-1}$, and the corresponding total energy
becomes   infinite.

This result completes  the proof of the stability of the solution
(3) against the axial perturbations with the finite value of the
initial energy.

\section{ The separation of the equations for $\psi_1(r)$ and $\psi_2(r)$ }

\hskip 14pt Now the question arises whether the system (14) can be split
into two
independent equations for some linear combinations of $\psi_1$,
$\psi'_1$, $\psi_2$ and $\psi'_2$. Let us obtain the general
condition on the coefficients $a$, $b$, $d$ and $\rho$ in (14) which
will permit one to say whether $2\times 2$ system can be split  or
not.

The weight $\rho$ can be eliminated from the equation (14) using the
 substitution $r\to\tilde r(r)$. Thus it appears to be   sufficient to
study the case $\rho=1$ only. With this restriction
 one will get the following equation:

$$ \psi''+\omega^2\psi+\pmatrix{d+a & b\cr b& d \cr}\psi =0. \eqno(17) $$

Let us suppose that there exists set of coefficients $\eta_1$,
$\eta_2$, $\eta_3$ and $\eta_4$ that the  linear combination

$$ \zeta = \eta_1\psi_1 +\eta_2\psi_2 + \eta_3\psi'_1 +\eta_4 \psi'_2
\eqno (18) $$

\noindent satisfies the following equation

$$ \zeta''+ \omega^2\zeta +\Omega\zeta=0 \eqno (19) $$

\noindent Note  that the coefficients in the substitution (18) should not
depend on $\omega$, otherwise the problem of the construction of
the coefficients $\eta_1$, $\eta_2$, $\eta_3$ and $\eta_4$  becomes
trivial. Moreover, the result obtained in this case appears to  be practically
useless. Indeed,  due to the "shadowing" produced by
the functions $\eta_1$, $\eta_2$, $\eta_3$ and $\eta_4$,  the behavior of
 function $\zeta$ after the substition (18)
 will not be directly connected with the behavior of the initial
function $\psi$.

By comparing the equations
(17)-(19), and separating the terms proportional to $\psi_1$,
$\psi_2$, $\psi'_1$, $\psi'_2$ and $\omega^2$, one can easily find  that

$$ \eta_3={\it const}, \qquad \eta_4={\it const}. \eqno(20) $$

\noindent It should be noted that the presence of the arbitrary constants
$\eta_3$ and $\eta_4$ corresponds to the orthogonal
rotation with the constant coefficients in the $(\psi_1,\psi_2)$
plane performed before the definition of the function $\zeta$ given by
(18).  Consequently,
keeping in mind the possibility of the preliminary constant
orthogonal rotation, we can choose $\eta_3=1$, $\eta_4=0$.
Then we can find explicit expressions for $\eta_1$ and $\eta_2$:

$$ \eta_1=-{1 \over 2} {b(r) \over \int b(r) dr} + {1 \over 2} \int a(r) dr
 - { \int \big(b(r) \int a(r) dr \big)  dr \over 2 \int b(r) dr },\eqno(21a) $$

$$ \eta_2={1 \over 2} \int  b(r) dr. \eqno(21b) $$

And, finally, the last equation of the system might be presented as:

$$a(r)+ 2d(r)+{\it const}_1={b' \over b}- {b \over 2\int b(r) dr} + $$
$$+ \big(\int b(r) dr\big)^2 +
{{\it const}_2 \over (\int b(r) dr)^2} +
{ \big(\int \big[ a(r) \int b(r) dr \big]dr\big)^2 \over 2
(\int b(r)dr)^2 }.\eqno(22) $$

\noindent This equation is the consistency condition.
It means that if it is fulfilled, then the coefficients $\eta_1$, $\eta_2$,
$\eta_3$ and $\eta_4$  given by (20)-(21)   satisfy  the
equations (17)-(19)  simultaneously.
By restoring the weight $\rho$ $(14e)$, we might obtain from (22)
the general form of the consistency condition as follows:

$$a(r)+ 2d(r)+{\it const}_1\cdot q^4(r) =  \eqno (23a)$$
$$={\sigma'' \over \sigma} - {1 \over 2}
\left( {\sigma' \over \sigma}\right)^2 + \rho^2\sigma^2 + {{\it
const}_2 \over \sigma^2} +
{ \big (\int  a(r) \sigma(r) dr\big)^2 \over 2 \sigma^2 }, $$

\noindent where the function  $\sigma(r)$ is  defined by

$$ \sigma(r) =- {1\over 2\sqrt {\rho(r)}} \int {b(r)dr \over
\sqrt{\rho(r)}}=- {1\over 2q(r)^2} \int {b(r)dr \over q^2(r)}.\eqno (23b) $$

Let us clarify the nature of the equation $(23a)$. It is well-known
that any equation of type (14) has a "dual" equation
\cite{Chandrasekhar}.  The simplest way to obtain the dual equation
is to rewrite the system (14) as the $4\times 4$ system of the
first-order equations:

$$ \left({\psi \atop \chi}\right)'= \pmatrix{ \hat M -  (q'/q)\hat E
 & i \omega q^2 \hat E \cr i \omega
q^2 \hat E & - \hat M -  (q'/q)\hat E  \cr}
\left({\psi \atop \chi}\right), \eqno (24) $$

\noindent where $\hat M$ is $2 \times 2$ matrix, $\hat E$ is
identity $2 \times 2$ matrix. Note that the matrix $\hat M$
always exists because it can be directly constructed as

$$ \hat M= \hat A'\hat A^{-1 }, \eqno (25a)$$

\noindent with the matrix $\hat{A}$ given by

$$ \hat A= \pmatrix{ \psi_1^{(1)} & \psi_1^{(2)} \cr \psi_2^{(1)} &
\psi_2^{(2)} \cr }.\eqno (25b) $$

\noindent where $\psi^{(1)}$ and $\psi^{(2)}$ are two linearly
independent solutions of the equation (14). The straightforward
calculation permits us to verify that the function $\psi$ from the
equation (24) satisfies   the system (14), and

$$ \pmatrix{ d+a & b \cr b & d \cr } = \left(\hat M - (q'/q) \hat E
\right)^2 + \left(\hat M - (q'/q) \hat E  \right)'.\eqno (26) $$

Similarly, the function $\chi$ is governed by the "dual" equation

$$ \left({ \chi_1 \atop \chi_2}\right)''+ \omega^2 q^4 \left({
\chi_1 \atop \chi_2}\right) + \pmatrix{\tilde d +\tilde a &\tilde b
\cr \tilde b& \tilde d \cr } \left({ \chi_1 \atop \chi_2}\right) = 0
,\eqno (27a)$$

\noindent where

$$ \pmatrix{ \tilde d + \tilde a  & \tilde b \cr \tilde b & \tilde
d \cr } = \left(\hat M + (q'/q) \hat E \right)^2 - \left(\hat M +
(q'/q) \hat E  \right)'.\eqno (27b) $$

One can verify  that  equation $(23a)$ is equivalent to the
condition $\tilde b=0$. It means that equations (17) can be separated only if
the "dual" system has a diagonal form.

The straightforward verification of the consistency of the equation
$(23a)$ for the values of $a$, $b$,$d$ and $\rho$ given by $(14b)-(14e)$
leads to rather complicated calculations. From the other side, by analyzing the
asymptotic behavior of condition $(23a)$ in the limit $r\to \mu$, we
might conclude that  (even taking into account the preliminary
constant ortogonal rotation in $(\psi_1,\psi_2)$ plane) equation
$(23a)$ can't be fulfilled in the limit $r\to\mu$. It leads us to
the conclusion that the system (14) is "essentially" two-dimensional
and can not be split  into two independent equations.

\section{Discussion.}

\hskip 10pt We have analyzed the problem of stability of the
exact solution of the standard
Einstein-Maxwell gravity coupled to an extra free massless scalar field.
It was shown  that, although the solution (3) contains naked singularities,
it is stable at least against   axial
perturbations. The problem of the stability of this solution
against polar perturbations is much harder to analyze. One unexpected
complication of these studies is that  the
differential operator for corresponding $3 \times 3$ eigenproblem
appears to be non-Hermitian.
However, this  research is currently in progress and the
obtained results will be reported in a subsequent publication.

Anticipating the possible questions we would like to note
that the correspondence of our analysis  to the
existing results on the perturbations of the Reisner-Nordstr\"om solution
is not quite strightforward. The reason for this   comes from the conclusion
that no smooth limit of our analysis exists for  $\phi\rightarrow 0$
and, although we can reduce our relations to the case $k=\pm1/2$,
the final results  will be degenerate at the point  $r=\mu$.
To show this, one may examine the case
$g_{00}/g_{11}\sim (r-\mu)^d$ with $d=2$ and  see that
the frequency term in the equation $(14a)$ is influencing  the asymptotic
behaviour of the function $\psi(r)$ at the vicinity of the surface $r=\mu$
\footnote{For the general case  with non-zero scalar field
($k\not= \pm1/2 \hbox{ and } |O| \not=\mu_0$), as we saw, $d<2$
and this influence is absent.}.
Moreover, the matrix structure of the  eigenproblem becomes:
$ \hat{E}\cdot a(r)+\hat{\sigma}_3\cdot b(r)\cdot const_1+
\hat{\sigma}_1\cdot b(r)\cdot const_2 $,
where $\hat{\sigma}_1$ and $\hat{\sigma}_3$ are  corresponding Dirac matricies.
As a result,  the equations for perturbations in the case of the
Reisner-Nordstr\"om solution    appear  to be split  into the following
three groups (reconstructing the already known result \cite{Chandra}):
(i) the scalar perturbations, which don't interact with other perturbations;
(ii) the two independent modes of the axial perturbations, where
both the gravity and the Maxwell field are mixed together, and (iii)
the two independent modes of the  polar  perturbations, which also contain
a mixture of  gravitational and electromagnetic fields.

Concluding this part, we would like to
note that the   stability of the general solution which
contains the naked singularities is well fit to the scenario  proposed
in the multi-dimentional extensions of the general relativity.
Thus, under the certain circumstances,
the Kaluza-Klein vacuum may decay by endlessly producing naked singularities.
This process from the five-dimensional point of view corresponds to Witten's
``bubbles of nothing'' which must eventually collide  \cite{Fay}, and so in
four-dimensions the singularities will coalesce.  However, it should be
emphasized that we have explored just the four-dimensional solutions
and the  further analysis of this problem should include the
non-trivial coupling of the scalar field in the higher dimensions.

The proven stability of the exterior static solution (3),  makes it
interesting to study whether  the scalar and electromagnetic
fields might be detected through the space gravitational  experiments.
Thus, following the standard procedure of the
$PPN$  formalism \cite{Nordtvedt}, one will find that only the  parameter
$\beta$
deviates from its general relativistic value, namely

$$\beta= 1+{\gamma \over c^2}{Q^2\over  2\mu_0^2}\eqno(28)$$

\noindent where $\mu_0$ is the Newtonian mass of the source
and we have restored the dimensional constants $\gamma$ and $c$.
This result  coincide with one for the Reissner-Nordstr\"om solution and
as long as  the metric (3) in post-Newtonian limit
 doesn't contains parameter $k$, the scalar field (defined by
the action (1)), can not be detected from the
data processing of the modern relativistic
 celestial mechanical experiments.

The second term in the expression  (28)
represents  the ratio of the electrostatic  energy
contribution in the
gravitational field produced by the same charged massive body. The
presence of this term  might
lead to an observable discrepancy in the motion of the celestial
bodies. For example, it will contribute to the Nordtvedt effect, which was
extensively studied in the Moon's motion \cite{Nordtvedt}.
 In the recent analysis of data obtained in Lunar
Laser Ranging which was carried out to detect the Nordtvedt effect,
a very tight  \cite{Dickey} limitation on the
parameterized post-Newtonian parameter $\beta$ was obtained:

$$\beta = 0.9999 \pm 0.0006   \eqno(29)$$

\noindent This result   suggests that   within the accuracy one
part in ten thousand, the contribution of the ratios of
electrostatic to self-graviational energies   presented by (29)
for    both   Moon and   Earth is negligable small.

Since gravity attracts positive and negative charges equally, then
the matter accreted on a massive astrophysical object will be nearly
neutral.  For the case of the celestial bodies with the gravitational energy
dominating over the electromagnetic one,
the parameter $\beta$ might be presented as:

$$\beta = 1 +  {\gamma_0\over c^2}{Q^2\over 4n^2 M^2_\odot}. \eqno(30)$$

\noindent where the mass of the star $\mu_0$  was expressed
in terms of the solar masses $M_\odot$: $\mu_0 = n M_\odot$.
The constraints imposed on new weak forces from the behavior of
the astrophysical objects gives for the maximum possible electric
charge $Q_{max}$ carried by celestial bodies the following
estimation:  $Q_{max} \le 10^{36} e$ \cite{Krause}. This
gives the following
estimation for the electrostatic energy contribution in the
parametrized post-Newtonian parameter $\beta$:

$$\Delta \beta = \beta - 1 \le { 2.17\over  n^2}\times 10^{-7}.
\eqno(31)$$

\noindent Unfortunately, even with $n= 1$  this result gives
practically unmeasureable value for the contribution of the
electrostatic energy of the charged astrophysical body to the
generated gravitational field.  According to this result,
the detection of the  electrostatic field contribution
  in the relativistic celestial mechanics experiments
performed in the weak gravitational field is presently impossible.

Thus, we have shown  that  the influence  of  both the electromagnetic and
the scalar fields
(given by the action (1)) on the motion of the asprophysical bodies
 is practically  unmeasureable in modern gravitational experiments.
However,  a wide class of  multi-dimensional
theories of gravity with an arbitrary number of
  massless scalar fields coupled to the usual tensor gravitational field
has been recently considered in \cite{Dam}. This investigation
was performed in order to analyze the
cosmological consequences of an inclusion of the multi-scalar field terms
 in the theory.
As a result, the authors of this paper
illustrated that although   these theories might
have coinciding post-Newtonian
limits with general relativity, they predict
non-Einsteinian behavior of the
stellar objects in a strong gravitational field.
In particular, it was noted  that this discrepancy will lead to
observable effects, for example,  for the binary pulsars.
This result makes it specifically interesting to
study the multi-scalar field extensions of the general
Einstein-Maxwell-scalar model together with the   condition
to  meet the experimental constraints based on the
tests of general relativity performed  to date.

PKS  was supported in part by the RFFI
Grant No 94-02-05490/91 and the St. Petersburg center for fundamental
research. SGT would like to thank the  National Research Council for
the support through the Resident Research Associateship award at the
Jet Propulsion Laboratory, California Institute of Technology. This work
was partially done in the Jet Propulsion Laboratory, California Institute
of Technology which is under  contract with the National Aeronautic and
Space Administration.

\end{document}